\documentclass[twocolumn,showpacs,preprintnumbers,amsmath,amssymb,epsf]{revtex4-1}

\usepackage{graphicx,epsfig}% Include figure files
\usepackage{dcolumn}% Align table columns on decimal point
\usepackage{bm}% bold math
\usepackage{subfigure}

\begin{document}

\title{Constraints on binary neutron star merger product from short GRB observations}
\author{He Gao$^{1}$, Bing Zhang$^{2,3,4}$, and Hou-Jun L\"{u}$^{5,6}$}
\affiliation{
$^1$Department of Astronomy, Beijing Normal University, Beijing 100875, China; gaohe@bnu.edu.cn
\\$^2$Department of Physics and Astronomy, University of Nevada Las Vegas, NV 89154, USA;\\
  $^3$Department of Astronomy, School of Physics, Peking University, Beijing 100871, China; \\
  $^4$Kavli Institute of Astronomy and Astrophysics, Peking University, Beijing 100871, China;\\
  $^5$GXU-NAOC Center for Astrophysics and Space Sciences, Department of Physics, Guangxi University, Nanning 530004, China\\
  $^6$Guangxi Key Laboratory for the Relativistic Astrophysics, Nanning 530004, China}

\begin{abstract}
Binary neutron star mergers are strong gravitational wave (GW) sources and the leading candidates to interpret short duration gamma-ray bursts (SGRBs).
Under the assumptions that SGRBs are produced by double neutron star mergers and that the X-ray plateau followed by a steep decay as observed in SGRB X-ray light curves  marks the collapse of a supra-massive neutron star to a black hole (BH), we use the statistical observational properties of {\em Swift} SGRBs and the mass distribution of Galactic double neutron star systems to place constraints on the neutron star equation of state (EoS) and the properties of the post-merger product.
We show that current observations already put following interesting constraints:
1) A neutron star EoS with a maximum mass close to a parameterization of $M_{\rm max} = 2.37\,M_\odot (1+1.58\times10^{-10} P^{-2.84})$ is favored; 2) The fractions for the several outcomes of NS-NS mergers are as follows: $\sim40\%$ prompt BHs,  $\sim30\%$ supra-massive NSs that collapse to BHs in a range of delay time scales, and $\sim30\%$ stable NSs that never collapse; 3) The initial spin of the newly born supra-massive NSs should be near the breakup limit ($P_i\sim1 {\rm ms}$), which is consistent with the merger scenario; 4) The surface magnetic field of the merger products is typically $\sim 10^{15}$ G; 5) The ellipticity of the supra-massive NSs is $\epsilon \sim (0.004 - 0.007)$, so that strong GW radiation is released post the merger; 6) Even though the initial spin energy of the merger product is similar, the final energy output of the merger product that goes into the electromagnetic channel varies in a wide range from several $10^{49}$ erg to several $10^{52}$ erg, since a good fraction of spin energy is either released in the form of GW or falls into the black hole as the supra-massive NS collapses.

\end{abstract}


\maketitle

\section{Introduction}
Mergers of two neutron stars (i.e. NS-NS mergers) are expected to be the primary source of gravitational waves (GWs) \cite{Merger} for upcoming ground-based interferometric detectors, such as the advanced LIGO, advanced VIRGO and KAGRA
interferometers \cite{GWdetector}. It has been proposed that NS-NS mergers could be associated with a variety of electromagnetic (EM) counterparts, detection of which would lead to direct confirmation of the astrophysical origin of the GW signals. The brightness of the EM counterparts depends on the details of the poorly known merger physics, especially the neutron star equation of state (EoS), and the outcome of the post-merger central remnant object \cite{EM1,EM2,gao15}.

Depending on the total mass of the NS-NS system and the NS EoS, NS-NS mergers could result in four different types of final products \cite{Paths,lasky14}: 1) immediate collapse into a black hole (BH); 2) a temporal hyper-massive NS (supported by differential rotation) which survives 10-100 ms before collapsing into a BH; 3) a supra-massive NS temporarily supported by rigid rotation, which collapses to a BH at a later time after the NS spins down; and 4) a stable NS. In this paper, we define the types 1) and 2) as ``prompt" BHs. The fractions for these outputs are currently not well constrained.

Short GRBs have long been proposed to originate from mergers of compact object binaries (NS-NS/BH mergers) \cite{SGRB}. Recent broad-band observational results, such as the mixed host galaxy types, non-detection of supernova associations, and relatively large offset of GRB locations in their host galaxies, lend support to such a suggestion \cite{berger14}. Thanks to {\em Swift}, the early X-ray afterglows of a large sample of SGRBs have been observed, which show rich features that demand extended engine activities, including extended emission \cite{EE}, X-ray flares \cite{Xflare}, and ``internal X-ray plateaus'' \cite{Rowlinson,Lu}.
In particular, an internal X-ray plateau is followed by a rapid decay which is too steep to be explained within the external shock model. It marks the abrupt cessation of the central engine, likely due to the collapse of a supra-massive NS into a BH \cite{Rowlinson,Lu}.

In this work, we make the assumptions that SGRBs are produced by NS-NS mergers, and that the internal plateaus are produced by a post-merger supra-massive NS, with the end of plateau marking the collapse time of the NS to a BH. We then use the available data (the observed fraction of SGRBs with internal plateaus, the distributions of the collapse time and plateau luminosity) to constrain NS EoS and the properties of merger product. Surprisingly, the available data already place interesting constraints on several parameters of the merger product.

\section{General formalism}
We consider an NS-NS system, in which the rest masses for the two NSs are $M_{\rm rest,1}$ and $M_{\rm rest,2}$, respectively. It is generally believed that NS-NS mergers conserve the rest mass, with $\lesssim 10^{-2}M_{\odot}$ of materials ejected during the merger \cite{Ejecta}, i.e. $M_{\rm rest,s}\approx M_{\rm rest,1}+M_{\rm rest,2}$, where $M_{\rm rest,s}$ is the rest mass of the nascent central merger remnant (henceforth, the central star). If the central star is a NS, its corresponding gravitational mass ($M_{\rm s}$) could be approximated as $M_{\rm rest,s}=M_{\rm s}+0.075M_{\rm s}^2$, where the masses are in units of the solar mass $M_\odot$ \cite{timmer96}.

Since before the merger the two NSs are in the Keplerian orbits, the central star should be rapidly spinning near the breakup limit. The rapid rotation could enhance the maximum gravitational mass ($M_{\rm max}$) allowed for NS surviving. For a given EoS, one may parameterize $M_{\rm max}$ as a function of the spin period of the central star \cite{lasky14},
\begin{eqnarray}
M_{\rm max} = M_{\rm TOV}(1+\alpha P^{\beta})
\label{Mt1}
\end{eqnarray}
where $M_{\rm TOV}$ is the maximum NS mass for a non-rotating NS, $\alpha$ and $\beta$ are functions of $M_{\rm TOV}$, star radius ($R$), and moment of inertia ($I$). Using the general relativistic NS equilibrium code {\tt RNS} \cite{stergioulas95}, the numerical values for $M_{\rm TOV}$, $R$, $I$ and thus $\alpha$ and $\beta$ for several EoSs have been worked out \cite{lasky14}.  In this work, we consider five EoSs derived in Ref. \cite{lasky14}, which
have a range of the maximum masses: EoS SLy \cite{SLy} with $M_{\rm TOV}=2.05\,M_\odot$, $R=9.99\,{\rm km}$, $I=1.91\times10^{45}\,{\rm g~cm^{-2}}$, $\alpha=1.60\times10^{-10}\,{\rm s}^{-\beta}$ and $\beta=-2.75$; EoS APR \cite{APR} with $M_{\rm TOV}=2.20\,M_\odot$, $R=10.0\,{\rm km}$, $I=2.13\times10^{45}\,{\rm g~cm^{-2}}$, $\alpha=0.303\times10^{-10}\,{\rm s}^{-\beta}$ and $\beta=-2.95$; EoS GM1 \cite{GM1} with $M_{\rm TOV}=2.37\,M_\odot$, $R=12.05\,{\rm km}$, $I=3.33\times10^{45}\,{\rm g~cm^{-2}}$, $\alpha=1.58\times10^{-10}\,{\rm s}^{-\beta}$ and $\beta=-2.84$; EoS AB-N \cite{AB} with $M_{\rm TOV}=2.67\,M_\odot$, $R=12.9\,{\rm km}$, $I=4.3\times10^{45}\,{\rm g~cm^{-2}}$, $\alpha=0.112\times10^{-10}\,{\rm s}^{-\beta}$ and $\beta=-3.22$; and EoS AB-L \cite{AB} with $M_{\rm TOV}=2.71\,M_\odot$, $R=13.7\,{\rm km}$, $I=4.7\times10^{45}\,{\rm g~cm^{-2}}$, $\alpha=2.92\times10^{-10}\,{\rm s}^{-\beta}$ and $\beta=-2.82$.
We note that the simple parameterization adopted in this paper following \cite{lasky14} could not fully catch the complicated physics of spin-dependent NS structure for a certain EoS and the variety of EoSs that are possible in nature. For example, in the parameterization formula, only $M_{\rm max}$ is explicitly presented. Another important parameter
characterizing an EOS, i.e. the radius of the most massive NS, is not specified, even though it would be implicitly represented by the $\alpha$ and $\beta$ parameters. Furthermore, there are many more EoSs discussed in the literature, which are not tested in this work. Our tests are relevant to the ensemble of EoSs that are close to these five representative EoSs.

By equating $M_{s}$ (the gravitational mass of the merger remnant) and $M_{\rm max}$ (the maximum NS mass taking rotation into
account), one can define a critical period
\begin{eqnarray}
P_c=\left(\frac{M_s-M_{\rm TOV}}{\alpha M_{\rm TOV}}\right)^{1/\beta}.
\label{Pc}
\end{eqnarray}
If $M_{s}$ is less than $M_{\rm TOV}$, the central star would settle to a NS that is uniformly rotating and eternally stable. On the contrary, if $M_{s}$ is greater than $M_{\rm TOV}$, the fate of the central star depends on the comparison between $P_c$ and the initial spin period $P_i$ of the central star. If $P_i$ is larger than $P_c$, rotation is not rapid enough and the central star would promptly collapse to a BH. If, however, $P_i$ is smaller than $P_c$, the effect of rotation is enough to support a NS, so that a long lasting rigid-rotation-supported supra-massive NS would survive for an extended period of time, which would collapse to a BH only when a good fraction of its rotational energy is lost and the centrifugal support can no longer support gravity.

For this last situation, the new born supra-massive NS would be initially differentially-rotating and entrained with strong magnetic fields. Within a timescale of $\sim0.1-1$ s \cite{shapiro00}, the combination of magnetic braking and viscosity would drive the star to the uniform rotation phase, during which we consider that the supra-massive NS losses rotation energy through both magnetic dipole radiation and GW emission, so that the spin-down law can be written as \cite{shapiro83}
\begin{eqnarray}
\dot{E}=I\Omega\dot{\Omega}=-\frac{32GI^2\epsilon^2\Omega^6}{5c^5}-\frac{B_p^2R^6\Omega^4}{6c^3},
\label{dotE}
\end{eqnarray}
where $\Omega=2\pi /P$ is the angular frequency and $\dot{\Omega}$ is its time derivative, $B_p$ is the dipolar field strength at the magnetic poles on the NS surface, $R$ is the radius of the NS, and $\epsilon$ is the ellipticity of the nascent NS. Define $a=\frac{32GI\epsilon^2}{5c^5}$, $b=\frac{B_p^2R^6}{6c^3I}$, and assume that $a$ and $b$ are approximately constant during the spin down of the NS, Equation (\ref{dotE}) can be solved analytically with initial condition $\Omega_i=2\pi/P_i$ and $\dot{\Omega}=\dot{\Omega}_i$ for $T=0$. The collapse time scale can be written as
\begin{eqnarray}
T_{\rm col}=\frac{a}{2b^2}\ln \left[\left(\frac{a\Omega_i^2+b}{a\Omega_{\rm col}^2+b}\right)\frac{\Omega_{\rm col}^2}{\Omega_i^2}\right]+\frac{\Omega_i^2-\Omega_{\rm col}^2}{2b\Omega_i^2\Omega_{\rm col}^2}.
\label{Tcol}
\end{eqnarray}
where
$\Omega_{\rm col}=2\pi/P_{c}$. At this time, a total amount of energy $E_{\rm rad}=1/2I(\Omega_{i}^2-\Omega_{\rm col}^2)$ has been released via both EM and GW losses, with the left spin energy collapsing into the BH. On average, the energy loss ratio between EM and GW can be written as
\begin{eqnarray}
\frac{\dot{E}_{\rm EM}}{\dot{E}_{\rm GW}}=\frac{\int_{\Omega_i}^{\Omega_{\rm col}} (b\Omega^{4})/(a\Omega^{6})d\Omega}{\int_{\Omega_{i}}^{\Omega_{\rm col}}d\Omega}=\frac{b}{a\Omega_{i}\Omega_{\rm col}}.
\label{Edotratio}
\end{eqnarray}
Even though the spin down solution $\Omega(T)$ includes both EM and GW losses, the observable total energy is in the EM channel, which can be estimated as
\begin{eqnarray}
E_{\rm EM,total}\simeq \frac{Ib(\Omega_{i}^2-\Omega_{\rm col}^2)}{2(a\Omega_{i}\Omega_{\rm col}+b)}.
\label{Edotratio}
\end{eqnarray}

At the collapse time of the supra-massive NS, the total luminosity in the EM channel is $L(T_{\rm col})=[B_p^2R^6\Omega_{\rm col}^4]/(6c^3)$. The luminosity of the internal plateau emission can be estimated as
\begin{eqnarray}
L_{b}=\frac{\eta B_p^2R^6\Omega_{\rm col}^4}{6c^3},
\label{Lbreak}
\end{eqnarray}
where $\eta$ is the efficiency of converting the dipole spin down luminosity to the observed X-ray luminosity.

\section{Observational constraints}
We analyze 96 SGRBs observed with {\em Swift} between 2005 January and 2015 October. The details of the data analysis method can be found in several previous papers \cite{Dataanalysis,Lu}. Basically, we extrapolate the BAT ($15-150$KeV) data to the XRT band ($0.3-10$ KeV) by assuming a single power-law spectrum \cite{BATtoXRT}, and then perform a temporal fit to the combined light curve with a smooth broken power law in the rest frame to identify a possible plateau (defined as a temporal segment with decay slope smaller than 0.5). In particular, we collect those bursts that exhibit a plateau followed by a decay index steeper than 3
as our ``internal plateau'' sample. The reason is that the steepest decay slope for the external shock model is $2+\hat{\beta}$ ($\hat{\beta}$ is the spectral index of flux density) due to the high-latitude ``curvature effect" emission from the relativistic jet \citep{Curvature}. A decay slope steeper than this value must then suggests an internal dissipation origin. Furthermore, since a $t^{-2}$ decay is expected by the magnetar dipole spin-down model \cite{shapiro83}, a decay slope much steeper than this would suggest a sudden cessation of energy injection, which is very likely due to the collapse of the supra-massive NS.
We find 21 candidates for supra-massive NSs, which comprises 22\% in the entire SGRB sample. This fraction should be regarded as a lower limit for supra-massive NSs formation fraction from NS-NS mergers, since 1) some of the SGRBs may be from NS+BH mergers; and 2) some supra-massive NSs may be missed from our selection if their collapse time is too late when the external shock emission becomes dominating \cite{Lu}.

Our first constraint comes from this fraction. We assume that the cosmological NS-NS merger systems have the same mass distribution as the observed Galactic NS-NS population.
Adopting a rest mass distribution of the merger product based on the Galactic population and applying the formalism discussed above, we can calculate the supra-massive NSs formation fraction as a function of the initial period $P_i$ for different EoSs. Figure 1 shows the curves for different EoSs, and only the GM1 EoS with $P_i\lesssim 1.2~{\rm ms}$ could reproduce the observed fraction. Under EoSs SLy and APR, most cases would immediately collapse into BH; wheares EoSs AB-N and AB-L would produce too many stable NSs. For EoS GM1, $P_i$ could not be much larger than 1.2 ms. Otherwise only a limited sample with $M_s$ very close to $M_{\rm TOV}$ could form a supra-massive NS. Our conclusion of favoring GM1 statistically is consistent with previous work \cite{lasky14,Lu}, which was based on case studies of individual SGRBs.

%%%%%%%%%%%%%%%%%%%%%%%
\begin{figure}[t!]%fig1
\vspace{0.3cm}
{\centering
\resizebox*{0.4\textwidth}{0.25\textheight}
{\includegraphics{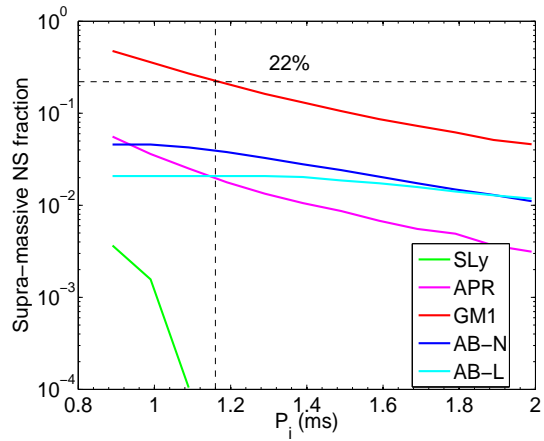}}
\par}
\caption{Supra-massive NS formation fraction as a function of initial period for different EOSs. The distribution of NS masses are generated following the observationally-derived distribution of Galactic NS-NS systems, i.e. $M_{\rm BNS}$ has a normal distribution $N(\mu_{\rm BNS}=1.32\,M_{\odot},\sigma_{\rm BNS}=0.11)$,
with a mean $\mu_{\rm BNS}$ and a standard deviation $\sigma_{\rm BNS}$ \cite{kiziltan13}.}
\label{spectrum}
\end{figure}
%%%%%%%%%%%%%%%%%%%%%%%%%%%

The minimum 22\% fraction of supra-massive NS population not only constrains the EoS, but also constrains the initial period $P_i$ of the merger product at birth. The fact that it has to be shorter than 1.2 ms is well consistent with the merger scenario, since the two pre-merger NSs are expected to be in Keplerian orbits. Assuming $P_i \sim 1$ ms (near break-up limit), and with the GM1 EoS, we predict the fractional distribution of the merger products is $\sim 40\%$ prompt BHs, $\sim 30\%$ supra-massive NSs, and $\sim 30\%$ stable NSs. Considering that the initial spin period may have a (narrow) distribution instead of a fixed value, these numbers would vary within a certain range, but the overall proportions should be close to our presented values, provided that the initial spin of the newly born supra-massive NS is near the breakup limit.

%%%%%%%%%%%%%%%%%%%%%%%%%%%
\begin{figure*}[t]
\subfigure[]{
    \label{fig:subfig:a} %% label for first subfigure
    \includegraphics[width=2.3in]{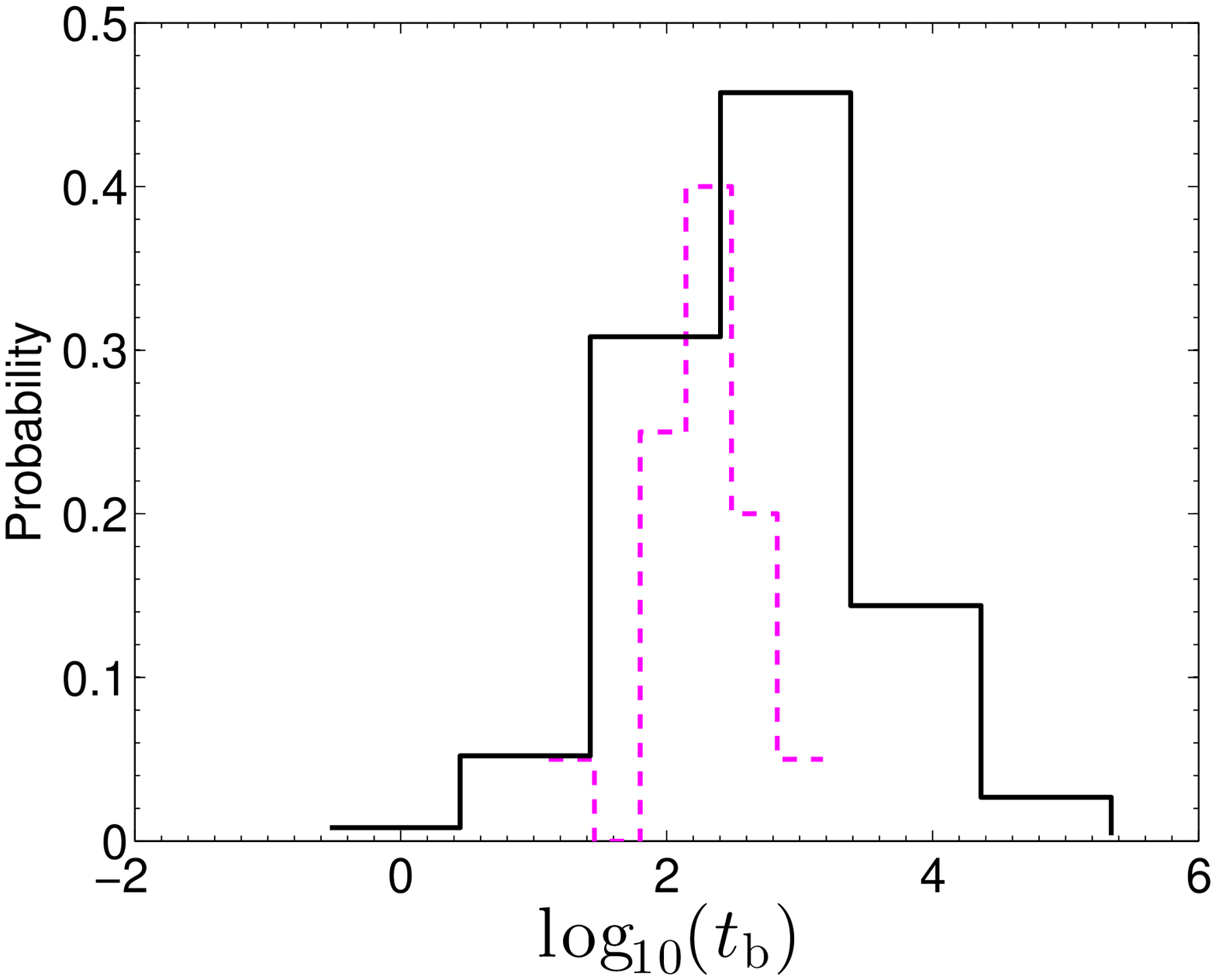}}
    \subfigure[]{
\label{fig:subfig:b} %% label for first subfigure
    \includegraphics[width=2.3in]{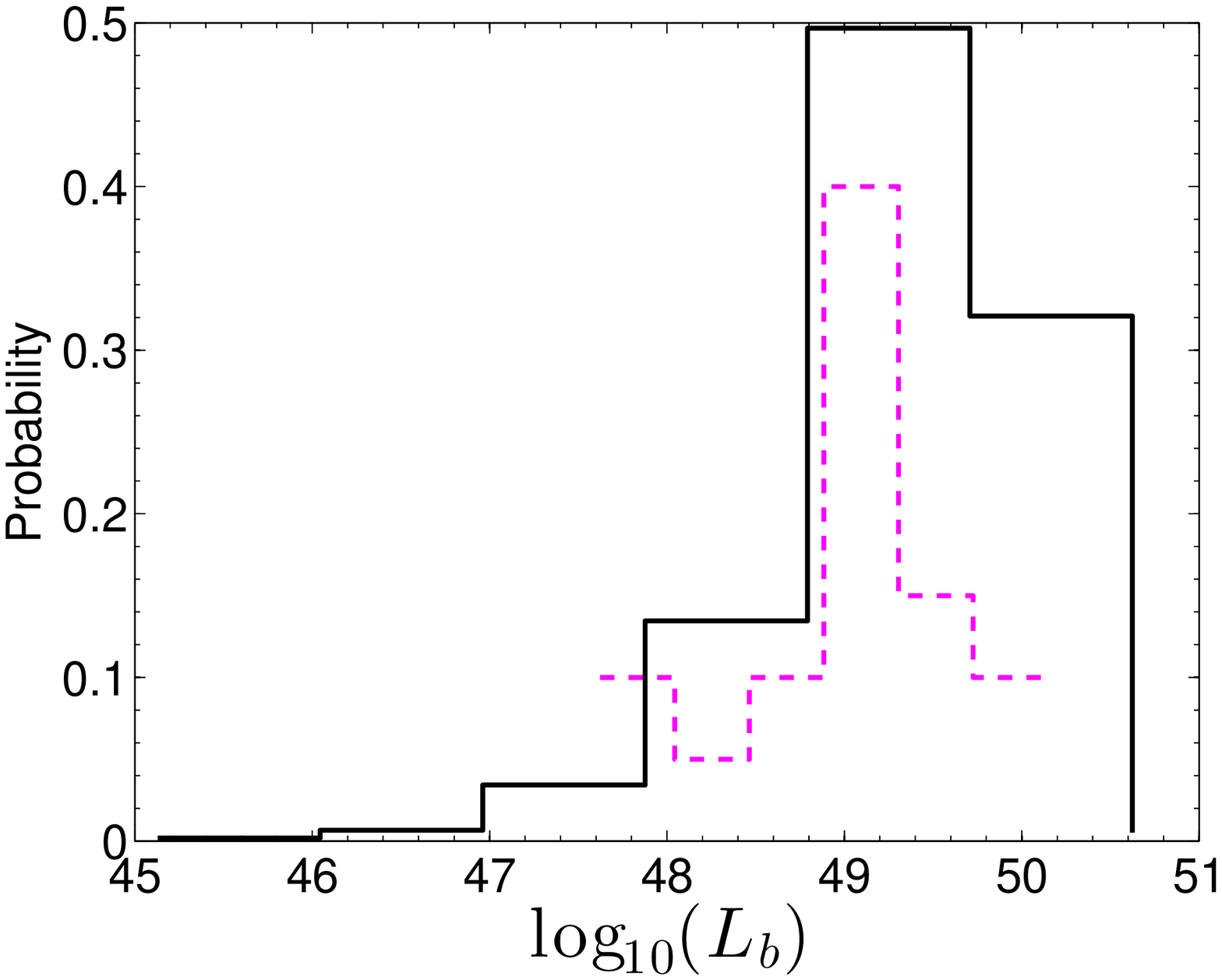}}
    \subfigure[]{
    \label{fig:subfig:c} %% label for first subfigure
    \includegraphics[width=2.3in]{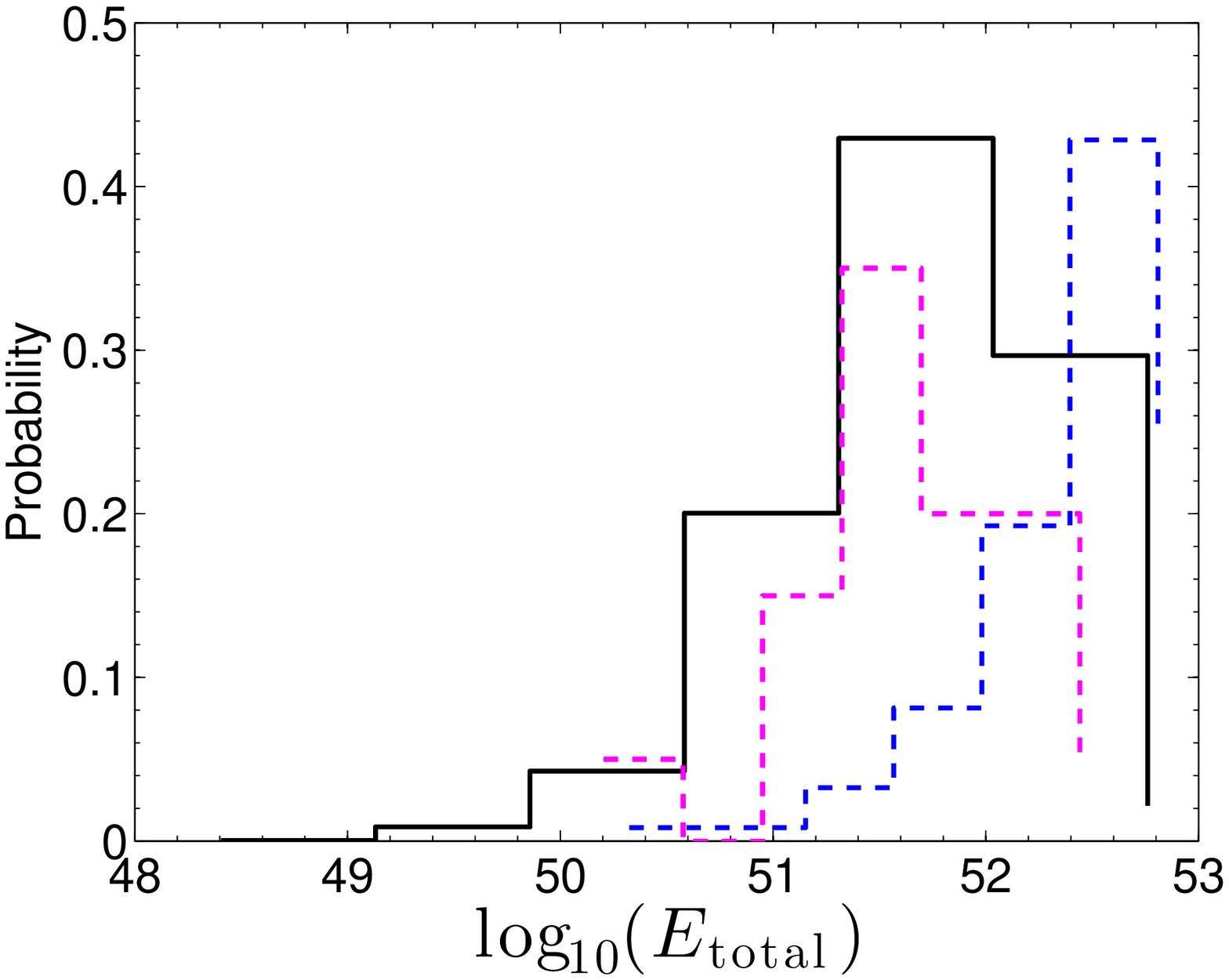}}
    \subfigure[]{
    \label{fig:subfig:c} %% label for first subfigure
    \includegraphics[width=2.3in]{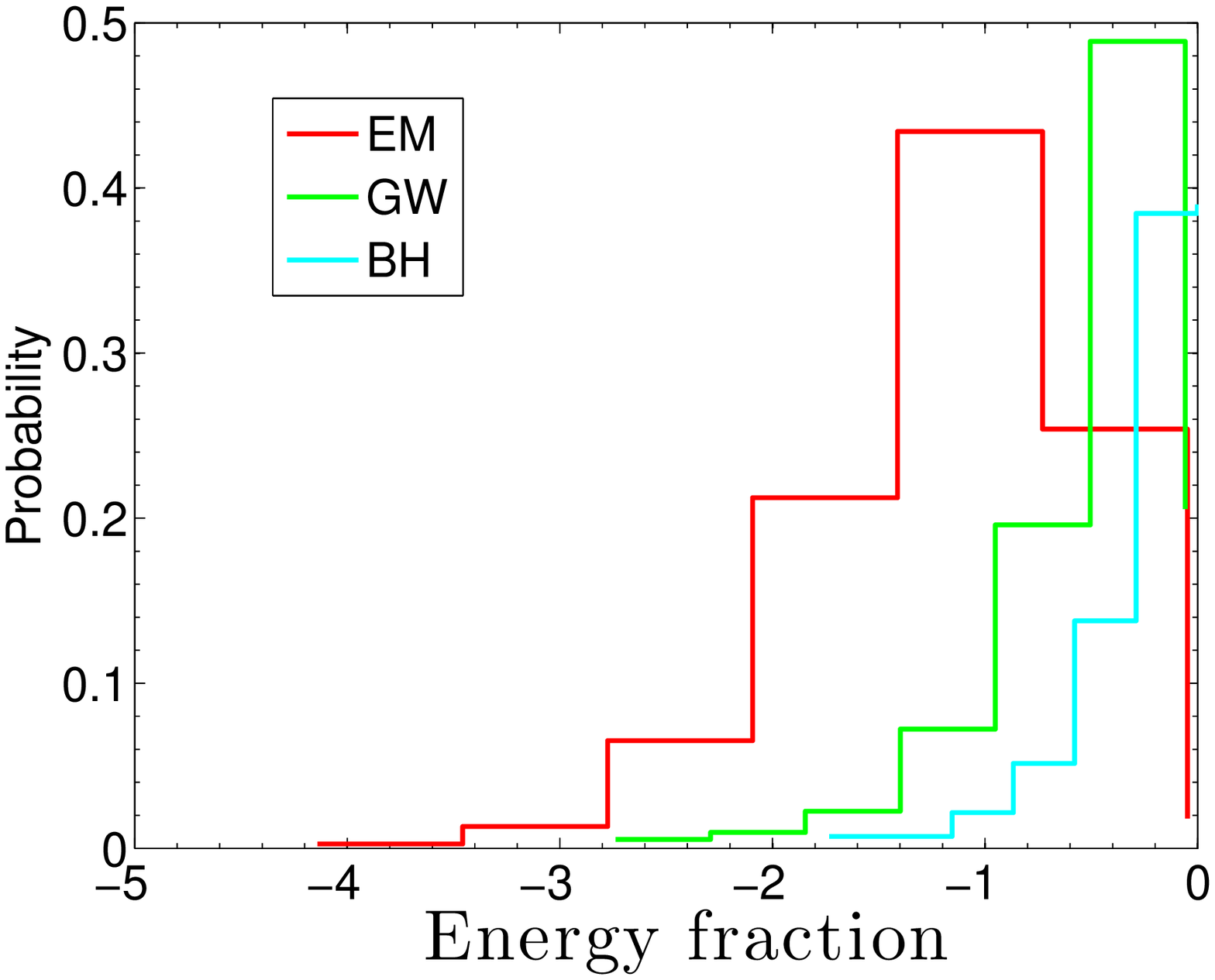}}
      \caption{Comparisons between the observed (pink dashed) and simulated (black) distributions for (a) the break time $t_b$ at the end of internal plateau, (b) the break time luminosity $L_b$, and (c) the total observed energy $E_{\rm total}$.  The parameters in the simulation are: $\epsilon=0.005$, $B_p\sim N(\mu_{\rm Bp}=10^{15}\,G,\sigma_{\rm BNS}=0.2)$ and $\eta=1$. The blue dashed line in panel (c) is an example when GW spin down is neglected. Panel (d) show the fractional energy distributions in the EM, GW, and BH channels.}
           \label{fig2}
            \end{figure*}
 %%%%%%%%%%%%%%%%%%%%%%%%%%%

Next, we focus on the supra-massive NS population (the internal plateau sample) and use three observed properties, i.e. the break time $t_b$ between the plateau and the steep decay which marks the collapse time of the supra-massive NS, the break time luminosity $L_b$, and the total observed energy in the EM channel $E_{\rm total}$, to constrain several other properties of the merger product.
The parameters $t_b$ and $L_b$ are directly collected from the observed light curves (shown as pink dashed histograms in Figure 2a and 2b). The total energy $E_{\rm total}$ could be estimated by summing the total isotropic radiation energy ($E_{\rm rad,iso}$) and kinetic energy ($E_{\rm K,iso}$). The former is derived in the rest-frame $1-10^4$ keV using the observed fluence through $k$ correction \cite{footnote,Lu}, and the latter is derived through afterglow modeling assuming a constant density media with standard parameters \cite{zhang07,density}.
The final distribution of the total released energy in the EM channel for the selected sample is shown as pink-dashed histogram in Figure 2c.

With the constrained EoS and $P_i$, the SGRB observational properties can be defined by three parameters of the supra-massive NSs: the dipolar field strength at the poles, $B_p$, the ellipticity of the neutron star, $\epsilon$, and the emission efficiency, $\eta$. Given $P_i$ and $P_c$, the collapse time distribution is determined by the spin-down law, which at different epochs may be dominated by either the EM or the GW energy loss term. If one turns off the GW spin down term,
one can reproduce the distributions of $t_{\rm b}$ and $L_{\rm b}$, but the distribution of $E_{\rm total}$ would completely deviate from the observations (blue dashed histogram in Figure 2c). This conclusion also holds for other EM spin down laws other than the dipole spin down law (e.g. \cite{siegel14}). The main reason is that without GW, most of the spin energy would have to be released in the EM channel, which for a millisecond rotator (spin energy is of the order of several $10^{52}$ erg), would greatly exceeds the observed energies of SGRBs. The results therefore demand significant GW radiation for nascent NSs \cite{fan13}. We then allow both GW and EM spin down and search for the parameter ranges that can simultaneously reproduce all three distributions ($t_{\rm b}$, $L_{\rm b}$, and $E_{\rm total}$). We search for the parameter regimes that can satisfy the three observational constraints. Interestingly, all three parameters, $B_p$, $\epsilon$, and $\eta$ can be reasonably constrained.
By requiring that the P values of the KS tests of all three distributions larger than 0.05 as the criteria for reproducing the observations, our Monte Carlo simulations suggest that the parameters are constrained in the following ranges: $\epsilon=0.004\sim0.007$, $B_p$ satisfies a normal distribution of $N(\mu_{\rm Bp}=10^{14.8-15.4}\,G, \sigma_{\rm BNS}\leq0.4)$, and $\eta=0.4-1$. The best P values for the KS tests of the three distributions reach 0.19, 0.97 and 0.45 for $t_{\rm b}$, $L_b$ and $E_{\rm tot}$, respectively, for  $\epsilon=0.005$, $B_p\sim N(\mu_{\rm Bp}=10^{15}\,G,\sigma_{\rm BNS}=0.2)$, and $\eta=1$ \cite{etafootnote}. The KS test is the poorest for the $t_{\rm b}$ distribution. However, the observed $t_{\rm b}$ distribution is likely subject to strong selection effect due to Swift XRT slew time and the first orbital gap, so that the true distribution is likely wider than what is observed and should be more consistent with the theoretical predictions.

Figure 2d shows the distributions of various energy components in merger products. Even though the initial spin energy is similar among different merger products, since a good fraction of the initial spin energy of the NS is released in the form of GW or falls into the BH as the supra-massive NS collapses, the total energy budget in the EM channel can vary in a wide range from several $10^{49}$ erg to several $10^{52}$ erg.

\section{Implications and discussion}
Assuming that SGRBs originate from NS-NS mergers, and that the internal plateaus are produced by supra-massive NSs that collapse at the end of the plateaus, we reached a series interesting constraints on the NS EoS and the properties of the merger product. The fractional distribution of the merger product is $\sim$40\% BH, $\sim$30\% supra-massive NSs, and $\sim$30\% stable NSs. The NSs are millisecond magnetars at birth, and the nascent NSs should release significant energy in the form of GWs. All these conclusions are either consistent with other observational constraints or with theoretical expectations. For example, the favored maximum NS mass $M_{\rm max} = 2.37\,M_\odot (1+1.58\times10^{-10} P^{-2.84})$ is well consistent with the observations of Galactic NSs and NS-NS binaries \cite{lattimer12}.
The required near-break-up $P_i \sim 1$ ms is fully consistent with the merger model.
 The high magnetic field strength $B_p \sim 10^{15}$ G is expected, since at such a fast spin, the $\alpha-\Omega$ dynamo mechanism likely operates \cite{dynamo}.

Observations demand a relatively large $\epsilon \sim (0.004-0.007)$ distribution. The origin of such a large $\epsilon$ is worth further investigation, but the inferred strong magnetic fields provide a natural source of distorting the NS and, hence, maintain a relatively large $\epsilon$ \cite{epsilon}. Within this scenario, in order to achieve $\epsilon \sim (0.004-0.007)$, a very high strength, $10^{16-17}$ G is needed, implying that the internal (toroidal) field may be more than $1-2$ orders of magnitude stronger that the surface value $B_p$ ($10^{15}$ G as constrained in this work. We note Ref. \cite{fan13} derived an even larger $\epsilon$ ($\sim 0.01$)  based on case studies of some SGRBs, and claimed that the GW may be detectable with the proposed Einstein Telescope \cite{ET}, but see \cite{lasky15}. Another comment is that $\epsilon$ likely evolves with time. Our Monte Carlo simulations suggest that a $50\%$ variation of $\epsilon$ is allowed without affecting our conclusions.

The fact that up to 60\% of the NS-NS merger products are either supra-massive or stable NSs
suggests that various EM counterparts \cite{EM2} of GW events due to NS-NS mergers are enhanced with respect to the case with a BH at the merger product \cite{EM1}. The collapse of a supra-massive NS to a BH may be also associated with a fast radio burst \cite{FRB}, the detection of which would lend support to the model discussed here.

In the end, we would like to point out several caveats of our approach. Firstly, we assume that the observed internal plateau in Swift SGRBs is produced by a post-merger supra-massive NS. Under this scenario, how the prompt SGRB emission is produced is still an open question. In the context of magnetar model for SGRBs, the prompt emission is assumed to be generated either by the drastic magnetic activity during the initial differential rotation phase \cite{gao06,rosswog07}, or by an accretion powered relativistic jet emerging from a NS-torus system shortly after the time of merger \cite{NS-disc}. However, a newly-born NS after a NS-NS merger would be accompanied with strong baryon pollution due to the dynamical ejecta from the merger and the neutrino wind from the hot merger product \cite{Ejecta,windejection}. Whether a clean relativistic jet may be launched remains unknown. A time-reversal scenario has been proposed to avoid baryon contamination \cite{time-reversal}, where the SGRB is generated after the supra-massive NS collapses into a BH with the jet powered by accretion. However, the formation of an accretion disk following the collapse of the supra-massive NS is questionable \cite{margalit15}. In this paper, we assume that a SGRB can be launched from the supra-massive NS shortly after the merger, and leave the launching mechanism as an open question for future investigations.

Secondly, we simply postulate that for all the SGRBs in our internal plateau sample, the end of plateau marks the collapse time of the NS to a BH. This assumption could be invalid if the observed X-ray light curve does not fully reflect the time evolution of the spin-down luminosity, for instance when the electromagnetic spin-down emission is strongly affected by the surrounding merger ejecta (Ref. \cite{gao15} for details). If this were the case, a late-time X-ray re-brightening signature would be predicted: as the ejecta becomes transparent and the magnetar wind would be still dissipating its energy and radiating X-ray photons. Considering that all the SGRBs in our internal plateau sample show no X-ray late re-brightening, we suggest that the rapid decay time is likely the neutron star collapse time for at least most cases discussed in our sample.

Finally, it is worth noting that the currently available data for SGRBs are still limited. Statistical errors may potentially bias our results. Moreover, the adopted method for distinguishing the internal/external plateau sample may also involve statistical uncertainties. The detailed numbers may differ, but the overall conclusions drawn in this paper would remain valid.

\begin{acknowledgments}
We thank the anonymous referees for constructive comments. This work is supported by National Basic Research Program (973 Program) of China (grants 2014CB845800), and the National Natural Science Foundation of China under grants 11543005. LHJ acknowledges support by Scientific Research Foundation of GuangXi University (Grant No XGZ150299). 
\end{acknowledgments}


\begin{thebibliography}{}
\bibitem{footnote}
We adopt the mean SGRB redshift $z=0.58$ If no redshift is measured.
\bibitem{density}
We tested two typical values for the medium number density $n=1~{\rm cm^{-3}}$ and $10^{-3}{\rm~cm^{-3}}$, and found that different $n$ values do not affect our conclusion much, and we chose $n=1~{\rm cm^{-3}}$ in our subsequent simulations. For SGRBs with a magnetar central engine, the outflow is likely not collimated. We therefore use the isotropic values of the total energy to be compared with the magnetar energy. See also \cite{Lu}.
\bibitem{etafootnote}
This best-fit value $\eta=1$ is likely too extreme. Our acceptable fits require $\eta > 0.4$, suggesting a high energy conversion efficiency.
\end{thebibliography}
\end{document}